\documentclass[prl,aps,preprint,superscriptaddress]{revtex4-1}
\usepackage{graphics}
\usepackage{epsfig}
\usepackage{amsmath}
\usepackage{amsfonts}
\usepackage{amssymb}
\usepackage{graphicx}
\usepackage{color}
\usepackage{tabularx}
\usepackage{txfonts}

\begin{document}

\title{High photon energy spectroscopy of NiO: experiment and theory}

\author{S. K. Panda}
\affiliation{Department of Physics and Astronomy, Uppsala University, Box 516, SE-751 20 Uppsala, Sweden}

\author{Banabir Pal}
\thanks{B.P. and S.K.P have contributed equally to this work.}
\affiliation{Solid State and Structural Chemistry Unit, Indian Institute of Science, Bangalore 560012, India}

\author{Suman Mandal}
\affiliation{Solid State and Structural Chemistry Unit, Indian Institute of Science, Bangalore 560012, India}

\author{Mihaela Gorgoi}
\affiliation{Helmholtz Zentrum Berlin für Materialien und Energie GmbH, Albert Einstein Stra\ss{}e. 15, 12489 Berlin, Germany}

\author{Shyamashis Das}
\affiliation{Solid State and Structural Chemistry Unit, Indian Institute of Science, Bangalore 560012, India}

\author{Indranil Sarkar}
\affiliation{Deutsches Elektronen-Synchrotron DESY, Notkestrasse 85, D-22607 Hamburg,
Germany}

\author{Wolfgang Drube}
\affiliation{Deutsches Elektronen-Synchrotron DESY, Notkestrasse 85, D-22607 Hamburg, Germany}

\author{Weiwei Sun}
\affiliation{Department of Physics and Astronomy, Uppsala University, Box 516, SE-751 20 Uppsala, Sweden}

\author{I. Di Marco}
\affiliation{Department of Physics and Astronomy, Uppsala University, Box 516, SE-751 20 Uppsala, Sweden}

\author{A. Delin}
\affiliation{Department of Physics and Astronomy, Uppsala University, Box 516, SE-751 20 Uppsala, Sweden}
\affiliation{Department of Materials and Nanophysics, School of Information and Communication Technology,
Electrum 229, Royal Institute of Technology (KTH), SE-16440 Kista, Sweden}
\affiliation{SeRC (Swedish e-Science Research Center), KTH, SE-10044 Stockholm, Sweden}

\author{Olof Karis}
\affiliation{Department of Physics and Astronomy, Uppsala University, Box 516, SE-751 20 Uppsala, Sweden}

\author{Y.O. Kvashnin}
\affiliation{Department of Physics and Astronomy, Uppsala University, Box 516, SE-751 20 Uppsala, Sweden}

\author{M. van Schilfgaarde}
\affiliation{Department of Physics, King's College London, Strand, London WC2R 2LS, United Kingdom}

\author{O. Eriksson}
\email{Electronic address: olle.eriksson@physics.uu.se}
\affiliation{Department of Physics and Astronomy, Uppsala University, Box 516, SE-751 20 Uppsala, Sweden}

\author{D. D. Sarma}
\altaffiliation{Also at Jawaharlal Nehru Centre for Advanced Scientific Research, Bangalore 560064, India}
\email {Electronic address: sarma@sscu.iisc.ernet.in}
\affiliation{Department of Physics and Astronomy, Uppsala University, Box 516, SE-751 20 Uppsala, Sweden}
\affiliation{Solid State and Structural Chemistry Unit, Indian Institute of Science, Bangalore 560012, India}
\affiliation{Council of Scientific and Industrial Research - Network of Institutes for Solar Energy (CSIR-NISE), New Delhi 110001, India}
\begin{abstract}
We have revisited the valence band electronic structure of NiO by means of hard x-ray photoemission spectroscopy (HAXPES) together with theoretical calculations using both the GW method and the local density approximation + dynamical mean-field theory (LDA+DMFT) approaches. The effective impurity problem in DMFT is solved through the exact diagonalization (ED) method. We show that the LDA+DMFT method alone cannot explain all the observed structures in the HAXPES spectra. GW corrections are required for the O bands and Ni-$s$ and $p$ derived states to properly position their binding energies. Our results establish that a combination of the GW and DMFT methods is necessary for correctly describing the electronic structure of NiO in a proper {\it ab-initio} framework. We also demonstrate that the inclusion of photoionization cross section is crucial to interpret the HAXPES spectra of NiO. We argue that our conclusions are general and that the here suggested approach is appropriate for any complex transition metal oxide. 
\end{abstract}

\maketitle

The electronic structure of the late transition metal monoxides (TMO)  has been studied intensely~\cite{Mott1937,PhysRevB.5.290,Brandow1977,PhysRevLett.52.1830,PhysRevB.30.4734,PhysRevLett.55.418,PhysRevB.44.943,
PhysRevB.54.13566,RevModPhys.70.1039,Dai2003,PhysRevLett.93.126406,PhysRevB.78.155112,PhysRevLett.100.206401,PhysRevB.79.235114,
PhysRevLett.103.036404,PhysRevB.82.045108,PhysRevB.91.245146}. Despite their apparent simplicity, they exhibit a rich variety of physical properties as regards magnetic, optical and transport properties. It is by now clear that most of these properties arise due to the electron correlations, in particular strong Coulomb interaction among the 3$d$ electrons of the transition metal ion. 
\par 
NiO is the archetype of TMO with strong correlation effects, and has often served as the system of choice when new experimental and theoretical methods are benchmarked. The electronic structure of NiO has remained enigmatic and controversial over the decades and have also become a major topic of text books on condensed matter physics~\cite{Mott_Orig,Brandow1977,PhysRevLett.52.1830,PhysRevB.30.4734,StefanHufner2003}. 
Over the years, a number of x-ray photoemission spectroscopy (XPS)  and bremsstrahlung isochromat spectroscopy (BIS) studies~\cite{PhysRevLett.34.395,PhysRevLett.53.2339,PhysRevB.33.4253,PhysRevLett.100.206401,Weinen2015} were carried out to address the electronic structure of this system. 
There also have been several theoretical attempts, ranging from model approaches~\cite{Brandow1977,PhysRevB.30.957,PhysRevB.33.8896,Sarma1990,PhysRevB.38.3449} to first-principles calculations~\cite{PhysRevLett.64.2442,PhysRevB.44.3604,PhysRevB.48.16929,PhysRevB.50.8257,PhysRevB.62.16392,PhysRevB.74.195114,
Takao2008,PhysRevLett.99.156404,PhysRevB.79.235114,PhysRevLett.109.186401,PhysRevB.91.115105} to explain the different spectroscopic manifestations of NiO. 
\par
Initially the electronic structure of NiO was interpreted using ligand field theory where the insulating gap is primarily determined by the large Coulomb interaction $U$ between Ni-$d$ states, an ideal case of a Mott insulator~\cite{Mott_Orig,Brandow1977}. 
This interpretation however could not be reconciled with resonance photoemission experiments~\cite{PhysRevB.26.4845,PhysRevB.53.10372}, a technique that provides insight into the orbital origin of valence band features. Specifically, ligand theory failed to capture the right character of the multi-electron satellite observed at high binding energy (around 9 eV). Later, Fujimori {\itshape{et al.}}~\cite{PhysRevB.30.957} suggested a cluster model which does a much a better job in explaining the experimental photoemission spectrum~\cite{PhysRevLett.53.2339}. In this picture, the high energy satellite arises due to the $d^7$ final state configuration which will be resonantly enhanced at the Ni 3$p$ threshold. This calculation was thus consistent with resonance photoemission results~\cite{PhysRevB.26.4845,PhysRevB.53.10372}. 
Based on their configuration interaction model,  Sawatzky and Allen~\cite{PhysRevLett.53.2339} also interpreted the XPS and BIS data of NiO and suggested that the fundamental gap opens between O-$p$ and Ni-$d$ states and is determined by $\Delta$, the so-called charge transfer energy.  
A few other studies~\cite{PhysRevB.44.3604,PhysRevLett.99.156404} also indicate that the first valence peak is originally a bound state coming from the strong hybridization of Ni 3$d$ and O 2$p$ states. 
This establishes NiO to be a charge transfer insulator, being, however, very close to an intermediate regime of the Zaanen-Sawatzky-Allen diagram where $U\approx\Delta$~\cite{PhysRevLett.55.418}. 
\par 
Although the cluster approach~\cite{PhysRevLett.53.2339,PhysRevB.30.957} with a configuration interaction model could explain most of the features in the experimental spectrum, it has a drawback in that it ignores the band aspects of all states, in particular the O 2$p$ states that play an important role as revealed from the angle-resolved photoemission spectroscopy (ARPES) experiments~\cite{Shen1989,PhysRevLett.64.2442,PhysRevB.44.3604}. Since band dispersion is a natural ingredient in first-principles based calculations of the electronic structure, there is a vast number of publications on NiO in this field~\cite{PhysRevLett.64.2442,PhysRevB.44.3604,PhysRevB.48.16929,PhysRevB.50.8257,PhysRevB.62.16392,PhysRevB.74.195114,
Takao2008,PhysRevLett.99.156404,PhysRevB.79.235114,PhysRevLett.109.186401,PhysRevB.91.115105}. However, the conventional effective single-particle band theory (e.g., within local (spin) density approximation -- L(S)DA) has been found to fail in most aspects of the transition metal oxides. For instance, it does in general not yield a band gap~\cite{PhysRevB.5.290,Cai2009} and for NiO this approach produces much smaller values of both the band gap and the local magnetic moment compared to experiment~\cite{PhysRevLett.53.2339}. The deficiency of the DFT/L(S)DA method was partially solved by adding an orbital dependent Hubbard $U$ term between the localized $d$ electrons in a mean-field fashion within the L(S)DA+ U approach~\cite{LSDAU,PhysRevB.44.943}. Although the LSDA+U method~\cite{LSDAU,PhysRevB.44.943} improves the values of the energy gap and local moment by a significant amount~\cite{PhysRevB.62.16392}, this theory still fails to provide an accurate description of the electronic excitation spectra.
Another theoretical approach applied to the transition metal oxides is the so-called GW approximation of Hedin~\cite{PhysRev.139.A796} which takes into account the dynamical screening of the long-ranged Coulomb interaction in a perturbative manner. However, when applied to NiO, the calculations based on GW method were found to not properly reproduce both XPS and BIS spectra at the same time~\cite{PhysRevLett.74.3221, PhysRevLett.93.126406}. This failure is most likely due to an insufficient treatment of the local correlation effects of the Ni $d$-orbitals, as was shown in Ref.~\onlinecite{PhysRevB.78.155112}. 
It is interesting to note that most of the studies on NiO were focused on its antiferromagnetic state, which is probably related to the well-known problems of GW to describe the paramagnetic (PM) phase of transition metal oxides~\cite{PhysRevB.78.155112}.
Moreover, in early theoretical works the presence of an insulating gap was even attributed to the existence of long-range magnetic order~\cite{PhysRevLett.52.1830}. 
However, it is known by now that NiO shows an insulating behavior even above the N\'eel temperature, i.e. in absence of antiferromagnetic
order~\cite{PhysRevB.54.10245}. This highlights the failure of any Slater-type formalism~\cite{PhysRev.82.538} in explaining the insulating behavior of NiO. 
More recent experiments~\cite{PhysRevB.54.10245,PhysRevB.70.195121} even suggest that long-range magnetic order has no significant influence on the valence band photoemission spectra and the electron density distributions. To overcome the limitations of the earlier methods, the electronic spectrum of the insulating PM phase of NiO  has also been calculated within the recently developed LDA+DMFT (dynamical mean field theory) scheme~\cite{PhysRevB.57.6884,RevModPhys.68.13} which offers a more sophisticated treatment of the correlation effects arising from Coulomb interactions within the Ni $3d$ levels. The LDA+DMFT approach successfully established that long-range AFM order has no significant influence on the valence band photoemission spectra and the large insulating gap in the PM phase is completely due to electron correlation effects~\cite{PhysRevB.74.195114,PhysRevLett.109.186401}. Furthermore, the valence band spectrum of NiO calculated within LDA+DMFT  agrees well with the experimentally measured spectrum recorded at 1.48 keV photon energy~\cite{PhysRevB.74.195114,PhysRevLett.109.186401}.
\par 
Recent experimental data~\cite{Weinen2015} show that the relative intensities of the various features of the NiO valence band photoemission spectrum are very sensitive to the incident photon energy. In particular, a new feature appear at higher binding energy ($\sim$ 7 eV) with increasing photon energy. These observations lack an explanation. Since photoemission spectra become less surface sensitive with an increase of the incident photon energy, these new experimental data must be viewed as a better representation of the valence band spectrum of bulk NiO, compared to previous experimental results. Hence, in order to claim a detailed understanding of the valence band spectrum of NiO, and transition metal oxides in general, one must theoretically reproduce the main features of these new experiments.
In view of the above, we report the experimental investigations of the valence band spectrum of NiO, to eliminate any doubts on the quality of these new observations~\cite{Weinen2015}, combined with theoretical calculations, that aim to provide an understanding of measured spectroscopic features. A detailed comparison of the theoretical results with the experimental ones reveals that a combination of LDA+DMFT and GW methods, together with a proper account of the cross sections for the photoemission process, are required for describing the experimental data. This establishes the minimal requirements for a complete theoretical framework for treating the valence band spectrum in a generic correlated oxide.  
\begin{figure}[t]
\includegraphics[width=0.99\columnwidth]{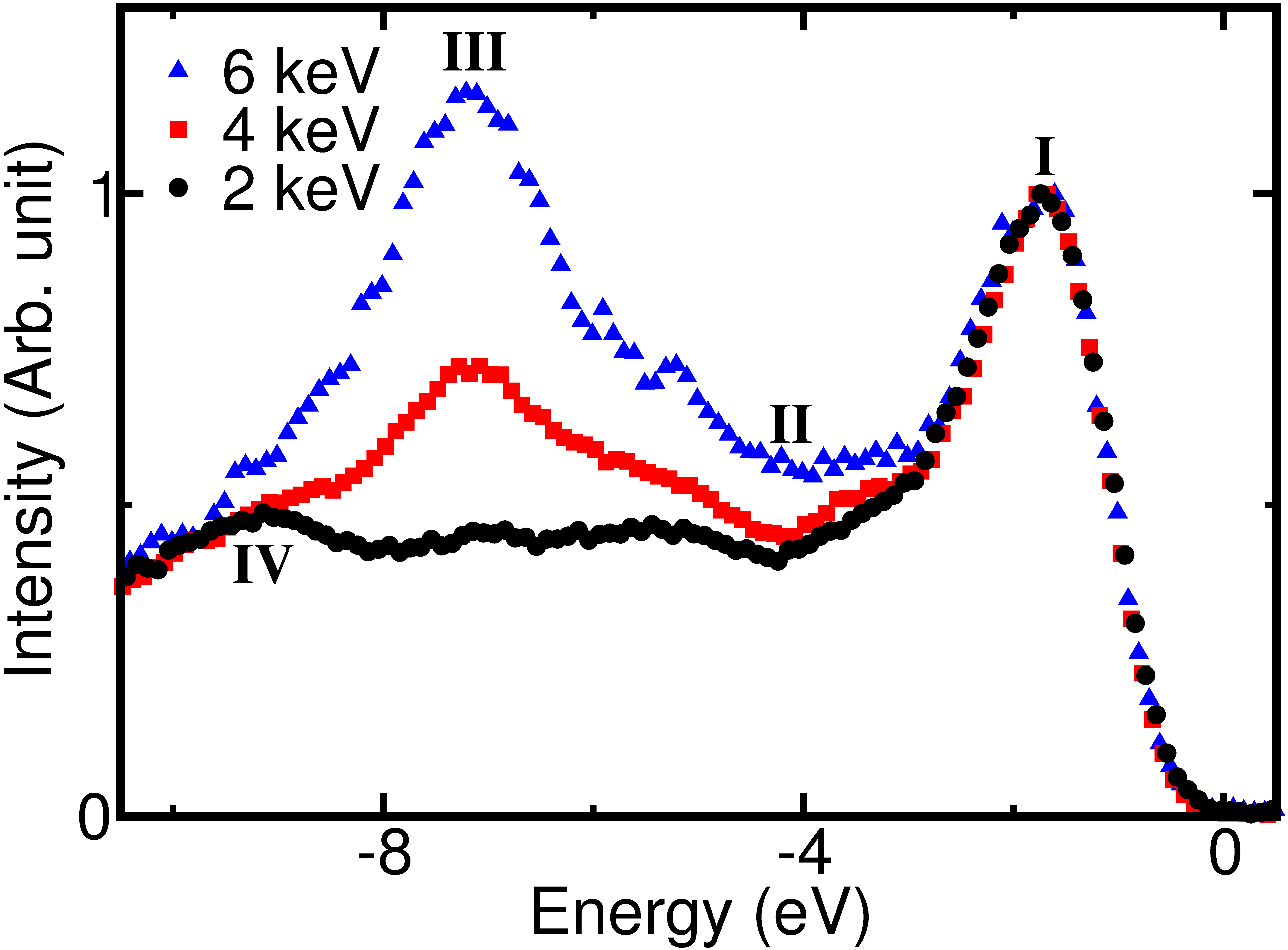}
\caption{HAXPES spectra of NiO valence band for three different photon energies (2 keV, 4 keV and 6 keV).}
\label{expXPS}
\end{figure}

\begin{figure*}
\includegraphics[width=0.99\columnwidth]{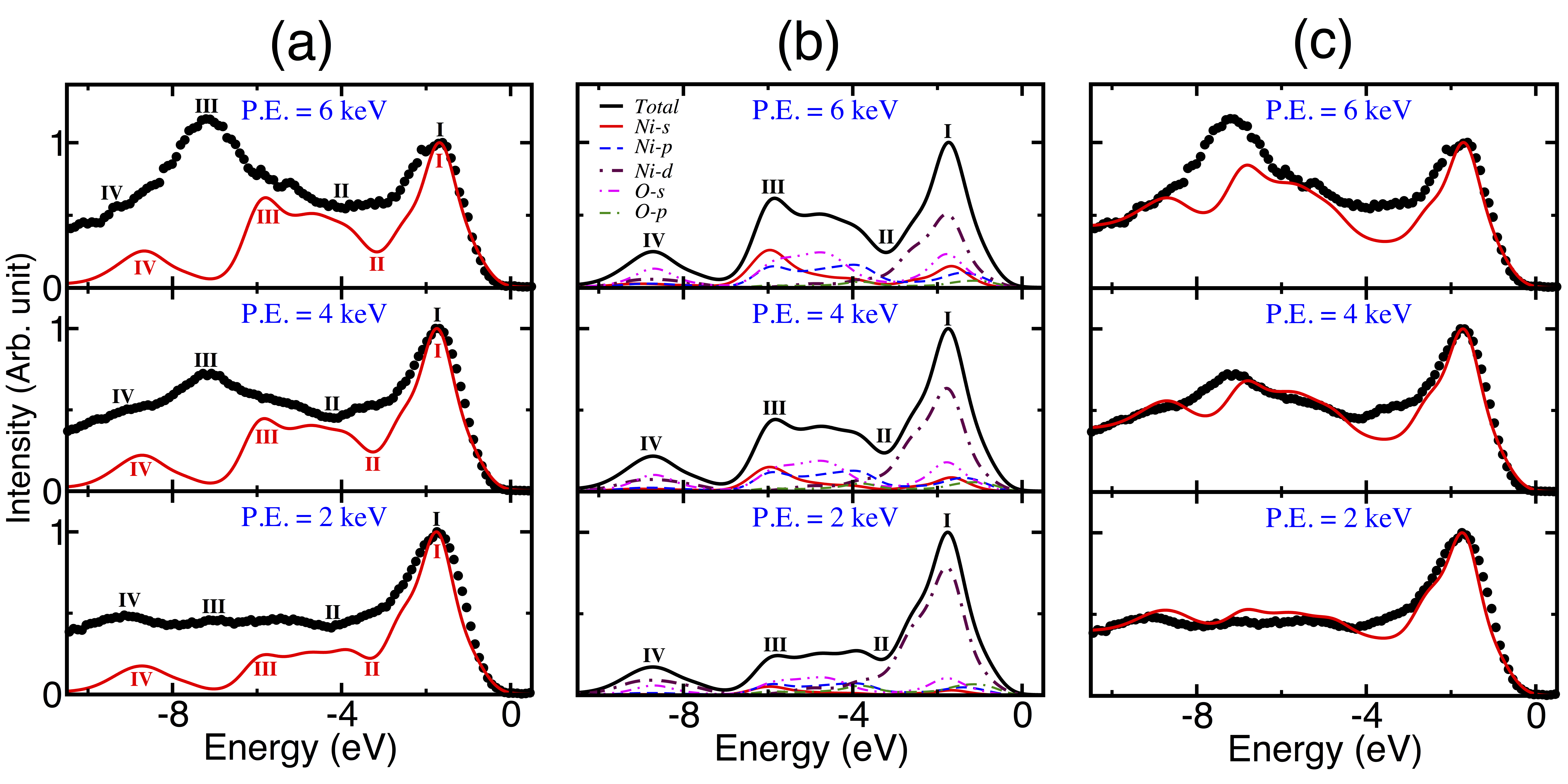}
\caption{(a) A comparison of our computed spectra (red line) from LDA+DMFT results with the measured HAXPES data (black circles). (b) Spectral intensities of all the Ni and O $l$ projected states to the total spectra are shown. (c) A comparison of our computed spectra (red line) after incorporating the GW corrections and the background effects with the measured HAXPES data (black circles). }
\label{thXPS}
\end{figure*}
\par 
The variable energy Hard X-ray photoelectron spectroscopy (HAXPES) experiments have been performed for three different photon energies (2 keV, 4 keV and 6 keV). Valence band spectra were collected at room temperature using a fixed, grazing incidence geometry ($\sim$ 10$^{\circ}$ grazing incidence). The base pressure in the HAXPES chamber was 1.0$\times$10$^{-9}$ Torr and the experiments were performed on a NiO single crystal freshly cleaved along the 100 direction before the measurements.
\par
Our theoretical investigation is based on the LDA+DMFT~\cite{PhysRevB.57.6884,RevModPhys.68.13} and GW~\cite{PhysRev.139.A796} methods, together with the calculation of energy-dependent matrix elements of the photoemission process~\cite{XPS1,XPS2}. 
Combining a sophisticated treatment of the many-body effects in NiO with an adequate description of the photoemission process will be shown to be necessary to explain how the experimental data depend on the energies of the incident photon. 
The calculations in LDA+DMFT were carried out in the paramagnetic phase by means of the RSPt code~\cite{FPLMTOCode}, based on the FP-LMTO method~\cite{FPLMTO_Orig,FPLMTO}. More details on this implementation can be found in Refs.~\onlinecite{DMFT_Rspt1,DMFT_Rspt2,DMFT_Rspt4}. The effective impurity problem for the Ni-3$d$ states was solved through the exact diagonalization (ED) method, as described in Ref.~\onlinecite{PhysRevLett.109.186401}. 
Strong Coulomb repulsion between Ni-$d$ orbitals was parametrised with the following values of Hubbard and exchange parameters: {$U$ = 8.5 eV} and {$J$ =0.8 eV}. 
A similar value of $U$ (8 eV) had also been used in an earlier DMFT study~\cite{PhysRevB.74.195114}. 
For the double counting correction, we used its fully localized limit formulation~\cite{PhysRevB.48.16929}. 
The calculations were carried out for a temperature of 300\,K. The fitting of the hybridization function for the ED simulation was done with two bath sites for each 3$d$ orbital. We have also carried out the quasi-particle self-consistent GW (QSGW) simulations following the implementation of Ref.~\onlinecite{PhysRevB.76.165106}. 
\par 
The photoemission spectra have been calculated within the single-scatterer final-state approximation~\cite{XPS1,XPS2}. Here the photocurrent is a sum of local (atomic-like) and partial ($l$-like) density of states (DOS) weighted by the corresponding cross sections. The cross sections are calculated using the muffin-tin part of the potential over the energy range of the corresponding DOS for the fixed photon energy. 
Overall, the cross sections calculated in this way depend on three quantities: the incident photon energy, the self-consistent potential and the binding energy of each state. The spectra has been broadened with a Gaussian to simulate the effect of spectrometer resolution. Finally a background function has been added in order to directly compare it with the experimental data. 
\par 
%
The results of the HAXPES measurement, on a freshly cleaved NiO (100) single crystal are shown in Fig.~\ref{expXPS}. The measurements were performed at room temperature at three different photon energies: 2 keV, 4 keV and 6 keV with the photoelectron momentum parallel to the polarization vector of the light. The valence band spectra of NiO at these three excitation energies are 
found to have several distinct features, which are marked in Fig.~\ref{expXPS}. Our measured spectrum at 2 keV shows that the most conspicuous peak (feature \textbf{I}) appears close to the fermi level, followed by a dip (feature \textbf{II}) and a small peak (feature \textbf{III}) around 7 eV binding energy. The multi-electron satellite (feature \textbf{IV}) is seen around 9 eV binding energy.  Interestingly, we observe that with increasing photon energy the spectral intensity of the feature \textbf{III} is significantly enhanced and becomes the dominant peak at 6 keV photon energy. We note that the earlier reported valence band XPS data~\cite{PhysRevLett.53.2339,StefanHufner2003} find a second Ni-$d$ peak just after the feature I at low binding energy. The absence of this structure in Fig.~\ref{expXPS} may be attributed to the larger broadening of the present experiment. Our experimental findings are consistent with previous reports~\cite{PhysRevLett.53.2339,Weinen2015}. 
\par 
As described above, we calculated the electronic structure of NiO within LDA+DMFT approaches using an ED implementation, which has earlier been successful to reproduce the spectroscopic data of NiO using Al K-$\alpha$ photon energy.~\cite{PhysRevLett.109.186401}. The results of the present calculations, displayed in Fig.~S1 of supplemental materials (SM), are in agreement with earlier DMFT studies~\cite{PhysRevB.74.195114,PhysRevLett.109.186401}. We would like to point out here that our calculations result in strong hybridization between Ni-$d$ and O-$p$ states, in combination with significant many-electron features of the Ni 3$d$ shell. In addition, we find substantial amount of Ni-$s$, O-$s$ and Ni-$p$ like states, located at valence energies that roughly coincide with feature \textbf{III} in Fig.~\ref{expXPS}.
\par 
Next, we computed the photoemission spectra as described above. 
In Fig.~\ref{thXPS}(a) we compare this calculation to our measurements. As we can see in Fig.~\ref{thXPS}(a), our computed spectra reproduce most features of the experimental HAXPES data for the three photon energies, both the peak at the top of the valence band, as well as the increased intensity of feature \textbf{III}, with increasing photon energy. The latter is found to be due to an increase of the cross section of the states around 6-7 eV binding energy, for the experiments which use a higher photon energy.
To illustrate this clearly, we present the spectral intensity of each $l$ projected state in Fig.~\ref{thXPS}(b). The contribution from O-$p$ states is very small throughout the entire energy range due to the low cross section of these states. Our calculations further show that the sharp peak near the Fermi level (feature \textbf{I}) has always predominant Ni-$d$ character for all probed photon energies. However the multi-electron satellite (feature \textbf{IV}) observed around 9 eV binding energy is not only due to the Ni-$d$ states, but has almost equal contribution from the O-$s$ states for the 2 keV and 4 keV photon energies. When we increase the photon energy to 6 keV, the O-$s$ contribution dominates (see Fig.~\ref{thXPS}(b)). The spectral intensity around feature \textbf{III} is found to originate from the cumulative contribution from Ni-$s$, Ni-$p$ and O-$s$ states for all photon energies. The most significant contribution to this peak is from the Ni-$s$ states whose relative cross section is higher in magnitude to the other states (primarily Ni-$d$) and increases rapidly with increasing photon energy. This gives rise to a sharp enhancement of the spectral weight of these states, at higher photon energies. 
\par 
It is however clear that the features \textbf{II} and \textbf{III} in our calculation are obtained at about 1 eV lower binding energy, compared to the experiment. Although the LDA+DMFT method can be applied to include local correlation effects for all orbitals at all sites, it does not include non-local fluctuations. The latter are usually quite significant for delocalized states of $sp$ character, also due to the fact that parametrizing the Coulomb interaction through local Hubbard terms may not be adequate.
The GW method is constructed to describe non-local correlation effects within fully {\it ab-initio} approach and without additional parameters. Its accuracy can be evidenced by comparing theoretical and experimental values of the band gap in $sp$ bonded systems~\cite{PhysRevLett.96.226402}. An already published GW study~\cite{PhysRevB.91.115105} of NiO indicates that the position of $s$ and $p$ derived states are shifted toward higher binding energy, compared to results obtained from LDA. To explore this as a possible explanation for the difference in binding energy of features \textbf{II} and \textbf{III} in Fig.~\ref{thXPS}, we carried out a quasiparticle fully self-consistent GW calculation~\cite{PhysRevB.76.165106}. This enables an estimate of the correct position and bandwidth of the $s$ and $p$ derived states. The results of our calculations are shown in Fig.~S2 of SM. A comparison of the QSGW and DMFT results as discussed in SM clearly reveals that QSGW pushes the $sp$ peaks around 6 eV towards higher binding energy, making the bandwidth larger by at least 1 eV compared to the DMFT bandwidth of those states. The size of these effects may appear large at first, on the same scale of the correlation effects for $d$-derived states. This is surprising, since traditionally it was always assumed that the Ni-3$d$ states are the crucial states to be localized. However, these results are not only the outcome of the most modern GW calculations, but also consistent with a recent study~\cite{ELF_CuO} of the electron localization function (ELF) in CuO which shows that the largest error due to the electron confinement in CuO is located at the O sites. We considered the renormalization of $sp$ states due to the GW corrections for the DMFT calculations. Hence we obtained an electronic structure that both contained effects of strong correlations of the $d$-shell as well as non-local correlations of $sp$-derived states. After multiplication with appropriate cross sections and adding a background contribution, we obtained the final spectra that are displayed in Fig.~\ref{thXPS}(c). The inclusion of background contribution was done to facilitate a direct comparison to experimental data that do contain a background due to secondary electrons. The agreement between the theoretical and experimental spectra shown in Fig.~\ref{thXPS}(c), is good for all photon energies. This applies to the positions of the different features, as well as their relative intensities. 
\par
Although a small disagreement in the intensity distribution for 6 keV photon energy is noticed, we observe that the intensity difference between the feature \textbf{II} and feature \textbf{III} is correctly obtained in our calculations when comparing to the corresponding experimental data. A more elaborate discussion to show the importance of matrix elements and GW corrections in order to interpret the experimental HAXPES data for each photon energy are provided in section III of the SM.  
\par 
In conclusion our detailed experimental and theoretical study reveals that the LDA+DMFT method as is widely used for understanding the electronic structure of strongly correlated systems, is not fully capable of interpreting the spectroscopic HAXPES data of NiO. GW corrections are important to properly position the binding energies of the delocalized $s$ and $p$ derived states, while DMFT properly describes the $d$ states. 
Our analysis indicates that a method combining GW and DMFT techniques~\cite{PhysRevLett.90.086402,SilkeDMFTGW}, is the method of choice in describing the electronic structure of NiO and most likely any complex transition metal oxide. Our results also show that the matrix element effects, which are often ignored, play a crucial role in understanding the electronic spectrum across a large range of photon energies. We have clearly shown that while effects from the photon energy dependent matrix elements are significant to obtain all the prominent features of the experimental spectra across the large range of photon energies, GW corrections are essential to position $sp$-derived features at correct binding energies. Therefore with the theoretical framework outlined in this manuscript, we can successfully reproduce all features of the valence band of NiO, including the high binding energy feature, that becomes most prominent at high photon energy. Thus the present study provides a very extensive analysis of the valence band electronic structure of NiO and most importantly suggests new routes for further theoretical analysis of the valence band spectrum of the transition metal oxides.


We acknowledge financial support from the Swedish Research Council (VR), Energimyndigheten (STEM), the Carl Tryggers Foundation (CTS), the Swedish Foundation for Strategic Research (SSF), the KAW foundation, eSSENCE and DST, India. 
The computer simulations were performed on resources provided by NSC and UPPMAX allocated by the Swedish National Infrastructure for Computing (SNIC). Valuable discussions with Diana Iusan, Barbara Brena and Jan Minar are acknowledged.



%
\end{document}


\title{Supplementary Materials for \\
High photon energy spectroscopy of NiO: experiment and theory}

\author{S. K. Panda}
\affiliation{Department of Physics and Astronomy, Uppsala University, Box 516, SE-751 20 Uppsala, Sweden}

\author{Banabir Pal}
\thanks{B.P. and S.K.P have contributed equally to this work.}
\affiliation{Solid State and Structural Chemistry Unit, Indian Institute of Science, Bangalore 560012, India}

\author{Suman Mandal}
\affiliation{Solid State and Structural Chemistry Unit, Indian Institute of Science, Bangalore 560012, India}

\author{Mihaela Gorgoi}
\affiliation{Helmholtz Zentrum Berlin für Materialien und Energie GmbH, Albert Einstein Stra\ss{}e. 15, 12489 Berlin, Germany}

\author{Shyamashis Das}
\affiliation{Solid State and Structural Chemistry Unit, Indian Institute of Science, Bangalore 560012, India}

\author{Indranil Sarkar}
\affiliation{Deutsches Elektronen-Synchrotron DESY, Notkestrasse 85, D-22607 Hamburg,
Germany}

\author{Wolfgang Drube}
\affiliation{Deutsches Elektronen-Synchrotron DESY, Notkestrasse 85, D-22607 Hamburg, Germany}

\author{Weiwei Sun}
\affiliation{Department of Physics and Astronomy, Uppsala University, Box 516, SE-751 20 Uppsala, Sweden}

\author{I. Di Marco}
\affiliation{Department of Physics and Astronomy, Uppsala University, Box 516, SE-751 20 Uppsala, Sweden}

\author{A. Delin}
\affiliation{Department of Physics and Astronomy, Uppsala University, Box 516, SE-751 20 Uppsala, Sweden}
\affiliation{Department of Materials and Nanophysics, School of Information and Communication Technology,
Electrum 229, Royal Institute of Technology (KTH), SE-16440 Kista, Sweden}
\affiliation{SeRC (Swedish e-Science Research Center), KTH, SE-10044 Stockholm, Sweden}

\author{Olof Karis}
\affiliation{Department of Physics and Astronomy, Uppsala University, Box 516, SE-751 20 Uppsala, Sweden}

\author{Y.O. Kvashnin}
\affiliation{Department of Physics and Astronomy, Uppsala University, Box 516, SE-751 20 Uppsala, Sweden}

\author{M. van Schilfgaarde}
\affiliation{Department of Physics, King's College London, Strand, London WC2R 2LS, United Kingdom}

\author{O. Eriksson}
\email{olle.eriksson@physics.uu.se}
\affiliation{Department of Physics and Astronomy, Uppsala University, Box 516, SE-751 20 Uppsala, Sweden}

\author{D. D. Sarma}
\altaffiliation{Also at Jawaharlal Nehru Centre for Advanced Scientific Research, Bangalore 560064, India}
\email {sarma@sscu.iisc.ernet.in}
\affiliation{Department of Physics and Astronomy, Uppsala University, Box 516, SE-751 20 Uppsala, Sweden}
\affiliation{Solid State and Structural Chemistry Unit, Indian Institute of Science, Bangalore 560012, India}
\affiliation{Council of Scientific and Industrial Research - Network of Institutes for Solar Energy (CSIR-NISE), New Delhi 110001, India}
\maketitle

\section{LDA+DMFT Electronic Structure}
\begin{figure}[h]
\includegraphics[width=0.55\columnwidth]{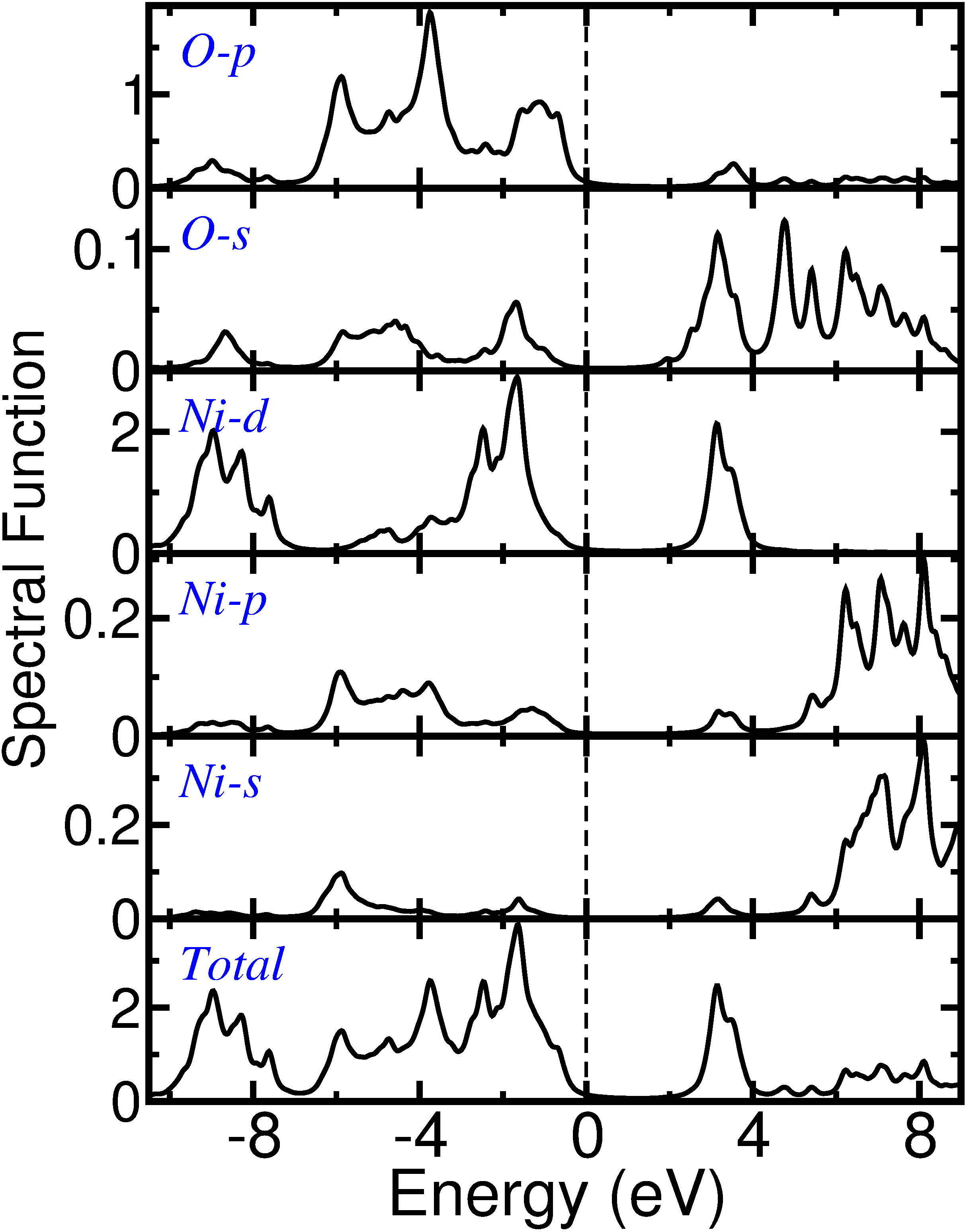}
\caption{Spectral functions of the Ni and O $l$ projected states.}
\label{spectratheory}
\end{figure}

We calculated the electronic structure of NiO within the LDA+DMFT approach using an an exact diagonalization (ED) solver. The k-integrated spectral functions of the $l$ projected states are displayed in Fig.~\ref{spectratheory}. As discussed in the main text, our results demonstrate a strong hybridization between Ni-$d$ and O-$p$ states, in combination with significant many-electron features of the Ni 3$d$ shell. The valence band maxima (VBM) have substantial admixture of Ni-$d$ and O-$p$ states. In addition, we find a substantial amount of Ni-$s$, O-$s$ and Ni-$p$ like states, located in the occupied part of the spectral function.
\section{QSGW Electronic Structure}
\begin{figure}[h]
\includegraphics[width=0.55\columnwidth]{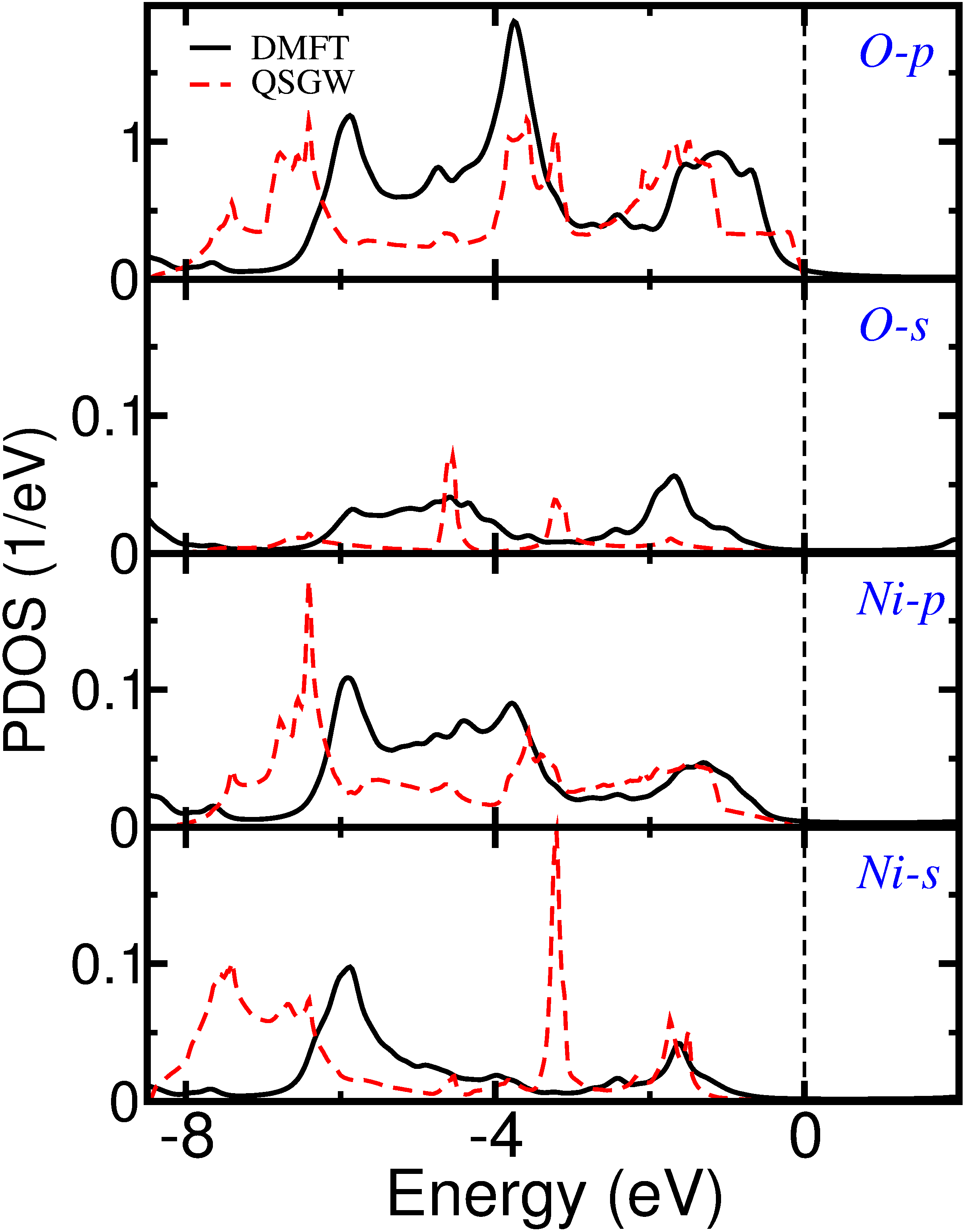}
\caption{The spectral functions of $s$ and $p$ states, obtained from DMFT and QSGW calculations.}
\label{spectraGW}
\end{figure}

As described in the main text, a comparison of the LDA+DMFT spectra with the experimental data indicate that the $s$ and $p$ derived states are positioned at lower binding energy compared to the experiment. It is known that the LDA+DMFT method can be applied to include local correlation effects for all orbitals at all sites, but it does include non-local fluctuations which is usually quite significant for the $sp$ states. In order to understand the influence of the non-local correlations on the $sp$ derived states, we have carried out quasi-particle self-consistent GW (QSGW) calculations. Such calculations are designed to describe the non-local correlations from first principles. The QSGW calculations are carried out in the spin-polarized phase in order to have an insulating solutions which is the true electronic state of NiO. The impact of the non-local correlation effects on the $sp$ derived states can be measured by comparing the results of our DMFT simulations with that of QSGW calculations. Such a comparison is displayed in Fig.~\ref{spectraGW}. The comparison reveals that the QSGW shifts the $sp$-states derived peak situated around 6 eV toward higher binding, compared to the DMFT results. Thus GW corrections lead to re-positioning of the peak III in the calculated spectra as has been discussed in the main text.     
\section{Theoretical photoemission spectra}
\begin{figure}[h]
\includegraphics[width=0.99\columnwidth]{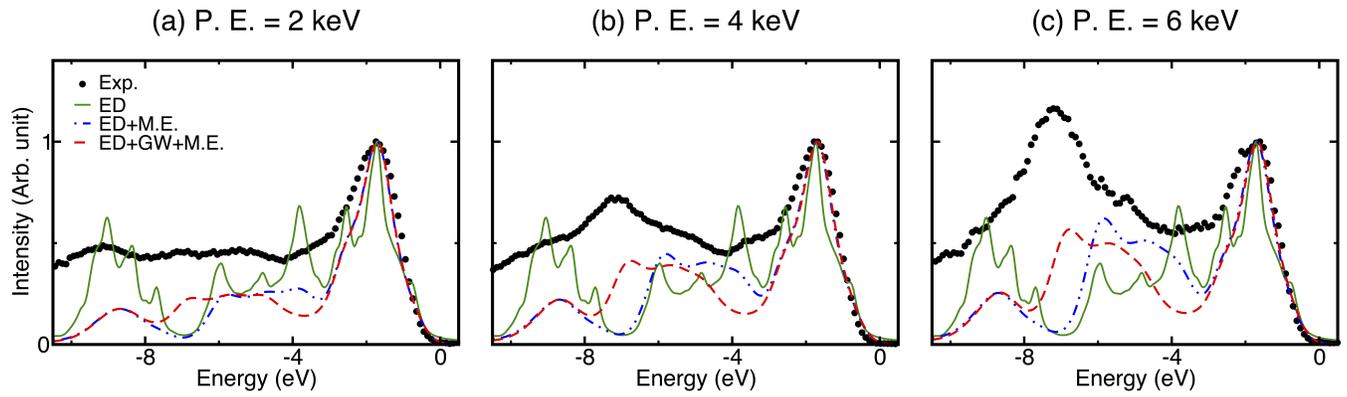}
\caption{A comparison of theoretical and experimental spectra to show the relative importance of various approaches.}
\label{PEStheory}
\end{figure}

Now we discuss the importance of various level of sophistication by comparing the theoretical and experimental photoemission spectra (PES) at three different photon energies. In Fig.~\ref{PEStheory}, we compare the experimental spectra (filled circles) with the normalized total spectral function, obtained from our ED simulation (green line), the PES obtained by including the matrix element effects for each $l$ states (dotted blue line) and the PES after including the GW corrections for $sp$ states and the matrix elements effects (dotted red line). Our results clearly reveal that the photon energy dependence of the matrix elements play a very crucial role to reproduce each and every features of the experimental spectra across the large range of photon energies and GW corrections are required to place those features at correct positions.   
